\documentclass[twoside]{article}

\usepackage{lipsum} 
\usepackage{aas_macros}

\usepackage[sc]{mathpazo} 
\usepackage[T1]{fontenc} 
\usepackage{microtype} 

\usepackage[hmarginratio=1:1,top=32mm,columnsep=20pt]{geometry} 
\usepackage{multicol} 
\usepackage[hang, small,labelfont=bf,up,textfont=it,up]{caption} 
\usepackage{booktabs} 
\usepackage{float} 
\usepackage{hyperref} 
\usepackage{natbib} 
\usepackage{graphicx}
\usepackage{caption} 
\usepackage{subcaption} 
\usepackage{lettrine} 
\usepackage{paralist} 
\usepackage{wasysym}
\usepackage{stmaryrd}

\usepackage{abstract} 

\usepackage{color}

\usepackage{ulem}

\usepackage{titlesec} 
\renewcommand\thesection{\Roman{section}} 
\renewcommand\thesubsection{\thesection.\arabic{subsection}} 
\titleformat{\section}[block]{\large\scshape\centering}{\thesection.}{1em}{} 
\titleformat{\subsection}[block]{\large}{\thesubsection.}{1em}{} 

\usepackage{fancyhdr} 
\usepackage{color}

\title{\vspace{-15mm}\fontsize{24pt}{10pt}\selectfont\textbf{Transmogrifiers: Bright of the Exomoon}} 

\author{
\large
\textsc{Michael B. Lund$^1$}\\
\normalsize $^1$Caltech/IPAC-NExScI\\
\normalsize \href{mailto:mlund@ipac.caltech.edu}{mlund@ipac.caltech.edu} 
\vspace{-5mm}
}
\date{}

\begin{document}

\maketitle 

\thispagestyle{fancy} 


\begin{abstract}

\noindent Though it may be a behavior that has been observed and documented for millennia, and despite the connection between it and the full moon, the astronomical community has afforded very little attention to lycanthropy. We hope to address this deficiency by using the population of known exoplanets as a natural experiment to better characterize what properties of the moon are necessary to trigger a transformation into a werewolf. We additionally investigate which exoplanets are most likely to have exomoons which may cause werewolves, with a particular focus on LHS 1140 b. We also propose a new mission called the Werewolves From Infrared Radiation and Spectral-typing Telescope, or WFIRST, in order to better characterize exoplanetary systems. This will allow us to explore the impact of stellar type on lycanthropy more than it has traditionally been considered. We believe this represents a major step forward in our understanding and recognition of the burgeoning field of exocryptozoology.

\end{abstract}


\begin{multicols}{2} 

\section{Introduction}
\lettrine[nindent=0em,lines=3]{S}ince time immemorial there have been stories of human transformation. The first story argued to be of werewolves date back over 4,000 years, to the Epic of Gilgamesh's story of a shepherd that is transformed into a wolf (leading to his unfortunate demise) (\citeauthor{Gilgamesh2100}, ca. 2100 BCE; \citeauthor{Kovacs1989}, \citeyear{Kovacs1989}), and the Greek story of Lycaon, who was also turned into a wolf and whose name has a connected etymology with the scientific term for the werewolf trait, lycanthropy \citep{Ovid8, Pausanias150}.

Though werewolves have long remained a part of humanity's lore, the connection between our lupine brethren and the moon would not come to light until much later. The first mention of the moon as "bright as day" in such a story, comes in the report of a man transforming into a werewolf in a graveyard at night (\citeauthor{Petronius90}, ca. 60). The first correlation between werewolves and the full moon (and could be argued to suggest causation) would follow a millennium later, when it was explicitly there were men that changed into wolves with the phases of the moon \citep{Gervase1211}. Still, it would take several hundred more years before the nature of werewolves would truly come to the attention of the astronomical community.

The first well-documented instance of modern astronomy crossing paths with lycanthropy didn't occur until the 20th century. Larry Talbot, an employee that worked on astronomical instrumentation at a company that specialized in optics that worked on astronomical instrumentation for Mount Wilson Obseservatory, first transformed into a werewolf while working on a private observatory in Wales. During his experience he was told of the moon's connection to werewolves \citep{Siodmak1941}:
\begin{quote}
Even a man who is pure in heart and says his prayers by night, may become a wolf when the wolfbane blooms and the autumn moon is bright.
\end{quote}
Decades later, there would be another intersection between astronomy and lycanthropy in the United Kingdom as werewolves would feature prominently on both sides of the Battle of the Astronomy Tower, which took place in Scotland on June 30, 1997 \citep{Rowling2014}.

There has been some work to suggest a link between werewolves and the moon, characterized not by moonlight but by the exchange of plasmas between the moon's exosphere and the Earth's atmosphere when the Moon passes into the tail of the Earth's magnetosphere \citep{Andrews2018}. While there is some extant discussion on the bridging of the atmospheres of a planet and its moon and how this impacts the life contained on the planet by allowing material to pass between the two atmospheres \citep{Veitch1993}, that explanation does not seem valid in cases where a werewolf reverts to human form due to cloudcover or other line-of-sight obstructions \citep{Ryan1988}. Instead, this behavior is more consistent with explanations that are linked to some sort of trigger associated with lunar light in particular \citep{Mattison2018}, but these mechanisms have not yet been tested in a controlled or systematic fashion. With the recent proposals and possible discoveries of exomoons, it is a very natural suggestion that the proliferation of exoplanets that may host exomoons could provide a natural laboratory to determine what factors are key for lycanthropy, even though exomoons have not yet been confirmed \citep{Heller2014}. There are some promising candidates, however these have generally been located around exoplanets that are not terrestrial, and so would not have a surface on which there could be werewolves \citep{Martinez2019, Kipping2022}, although there is some work that has looked at werewolves in space in the past \citep{Reeves-Stevens1995}.

In this paper, we examine the population of known exoplanets in order to identify how many exoplanets are suitable candidates for hosting a moon that would sufficiently illuminate the planet's surface in order to facilitate transformation. We begin by collecting our data and conducting our analysis in Section~\ref{Methods}. We assess the implications of these results and discuss further steps in in Section~\ref{Discussion} and propose a new mission called the Werewolves From Infrared Radiation and Spectral-typing Telescope, or WFIRST, in order to better characterize exoplanet systems for werewolf habitability. We also discuss several of the new directions in which further research can proceed. Finally we summarize this work in Section~\ref{Summary}.

\section{Methods} \label{Methods}

\subsection{Data} \label{Data}
For gathering our exoplanet data, we use the tables available at the NASA Exoplanet Archive\footnote{https://exoplanetarchive.ipac.caltech.edu/}, supported by the NASA Exoplanet Science Institute, which compile all published and confirmed exoplanets and make them easily accessible via a Table Access Protocol (TAP). We use the data available from the Planetary Systems (PS) Table \citep{PST}, as this provides internally consistent parameters. At present, this table is comprised of over 5,000 exoplanets, although not all exoplanets have complete sets of published data or are likely to be terrestrial planets and so some curation is required.  

Of the initial 5,005 exoplanets, we begin by removing any planets that are missing values for any of the following properties for the planet or host star: stellar luminosity, stellar mass, stellar radius, planetary mass, planetary radius, planetary density, and planetary semi-major axis; any planets missing orbital eccentricity data are treated as having circular orbits with an eccentricity of 0. This cut leaves us with 301 planets. Next, we constrain our focus to terrestrial exoplanets by removing all exoplanets that are likely to be gaseous and keeping those that are likely to be rocky. We do this by utilizing work suggesting that planet radius is a proxy for composition, and that this boundary exists at 1.75 $R_{\oplus}$, leaving us with 38 planets \citep{Lopez2014}. We also apply a cut based on mass, where the distribution of masses for rocky exoplanets ends at around 25 $M_{\oplus}$, and so again any planets with masses larger than this are removed, leaving us with 38 vetted (and likely terrestrial) planets \citep{Otegi2020}.

In Section~\ref{Analysis}, we will also briefly look at what sort of results we get when looking at the larger Planetary Systems Composite Parameters (PSCP) Table \cite{PSCT}, where individual planets do not necessarily have self-consistent parameters, but for the bulk of this work, we will focus our attention on this initially-defined subset.

\subsection{Calculations} \label{Calculations}

The two key planetary and system parameters that impact exomoons we begin our investigation with are the Hill sphere and the Roche limit. The Hill sphere represents the boundary between where the gravity from a smaller body (our exoplanet) will be stronger than the gravity from a larger body (the host star). Any exomoons would need to be contained within the Hill sphere, representing an upper orbital radius boundary for any possible exomoons. Following the work in \citet{Hamilton1992}, for an orbit with some eccentricity $e$ of an exoplanet with mass $m$ and a semi-major axis $a$ around a host star of mass $M$, we can express the radius of the Hill sphere $r_{H}$ as follows: 
\begin{equation}
    r_{H} \approx a (1-e) \sqrt[3]{\frac{m}{3M}}
\end{equation}

We can similarly bound how close the exomoon can get to the exoplanet by determining at what distance the tidal forces from the exoplanet will exceed the self-gravitation of the moon by determining the Roche limit \citep{Roche1849, Shu1981}. For two rigid bodies, the Roche limit $r_{R}$ can be expressed as a function of a function of the primary object's mass $M$ and the secondary object's mass $m$ and radius $R_{m}$:
\begin{equation}
    r_{R} = R_{m}\sqrt[3]{\frac{2M}{m}}
\end{equation}
For our purposes, however, it will be more useful to use an equivalent form of this equation that is written in terms of the primary object's radius $R_{M}$ and density $\rho_{M}$ and the secondary object's density $\rho_{m}$:
\begin{equation}
    r_{R} = R_{M}\sqrt[3]{\frac{2\rho_{M}}{\rho_{m}}}
\end{equation}
As we are looking for exomoons that will behave analogously to our own moon, we set the secondary density to be equivalent to the mean density of our own moon, $\rho_{\leftmoon} = 3.344 g/cm^{3}$, as stated in reference material\footnote{Sourced from Wolfman Alpha's natural language search}. This simplifies our equation so that it may be expressed in terms of the already-characterized exoplanet:
\begin{equation}
    r_{R} = R_{M}\sqrt[3]{\frac{2\rho_{M}}{\rho_{\leftmoon}}}
\end{equation}

We then have two approaches for placing moon-analogues around these exoplanets. The first is to place a Moon-sized exomoon at the same distance as our own moon, 385,000 km \citep{Chapront-Touze1988}. In the second case we consider the long-standing observation that the sun and moon appear to be the same size in the sky (\citeauthor{Aristarchus270}, ca. 270 BCE; \citeauthor{Heath1913}, \citeyear{Heath1913}), and the relative paucity of werewolves elsewhere in our solar system, as an indication that this unique ratio is important. We then use this shared apparent size to determine the distance at which the the moon must orbit its host planet. This distance $d_{m}$ can be expressed in terms of the exomoon's radius $R_{m}$ and the exoplanet's semi-major axis and the host star's radius $R_{\star}$:
\begin{equation}
    d_{m} = \frac{a R_{m}}{R_{\star}}
\end{equation}
Here we explore our two possibilities for the exomoon's radius, one where we hold it to be at one lunar radius and the other where we hold it to have the same ratio to the planet as the earth has to to the moon. We also require in all cases that the exomoon must be outside the Roche limit but inside the Hill sphere, but we do not take any other dynamical effects into consideration to determine stability, including the impact of tightly packed planetary systems.

In all cases, our equation for calculating out the flux from the exomoon will then be the following function of albedo $\alpha$, stellar luminosity $L_{\star}$, planetary semi-major axis a, exomoon radius $R_{m}$, and exomoon semi-major axis $d_{m}$:
\begin{equation}
    F_{m} = \frac{\alpha L_{\star}}{4\pi(a+d_{m})^{2}}(\frac{R_{m}}{d_{m}})^{2}
\end{equation}

We can compare this to the fluxes in our own solar system, in terms of solar luminosity $L_{\odot}$, Earth's semi-major axis, a distance from the earth to the moon of 385,000 km \citep{Chapront-Touze1988}, a radius of the moon of 0.2727 $R_{\oplus}$, and albedo of 0.12:
\begin{equation}
    F_{\star} = \frac{L_{\odot}}{4\pi(1 AU)^{2}}
\end{equation}
\begin{equation}
    F_{\leftmoon} = \frac{0.12 L_{\odot}}{4\pi(1 AU+0.27 R_{\oplus})^{2}}(\frac{0.27 R_{\oplus}}{385000 km})^{2}
\end{equation}
This gives us our values for the Earth of a $F_{\star}$ of $1361 W/m^{2}$ and a $F_{\leftmoon}$ of $3.32*10^{-3} W/m^{2}$, or a difference of about 14 magnitudes. Our solar system measurement for $F_{\leftmoon}$ will be used as our threshold for determining if there will enough flux from an exomoon to transform a human to a werewolf.

\subsection{Analysis} \label{Analysis}
\begin{figure*}[!htb]
  \begin{center}
   \includegraphics[width=0.70\textwidth]{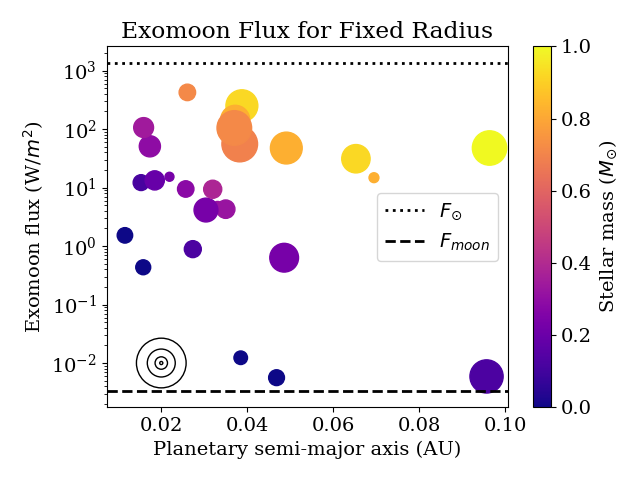}
   \includegraphics[width=0.70\textwidth]{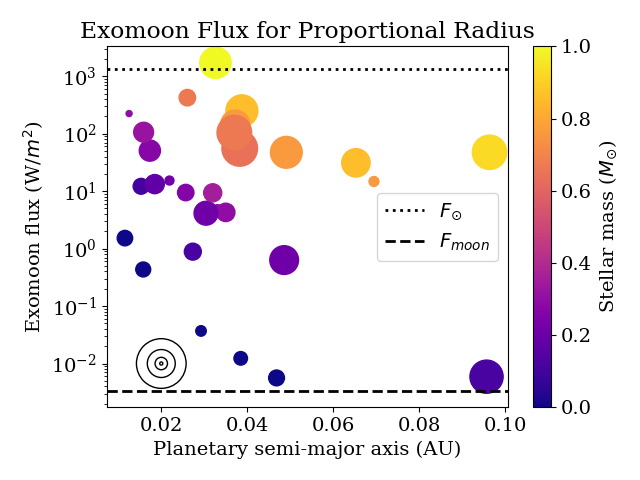}
  \end{center}
  \caption{The exomoon fluxes calculated for the vetted exoplanets from the Planetary Systems table, with the planetary semi-major axis on the x-axis and the exomoon flux received by the planet on the y-axis. Points are scaled as a function of planetary radius, and the concentric circles in the bottom left represent radii of 0.5, 1, 1.5, and 2 Earth radii. Colors correspond to the host star's mass. The horizontal dashed line marks the flux the Earth receives from the full moon and the horizontal dotted line represents the flux the Earth receives from the Sun. The top figure is for a lunar-radius moon placed at the appropriate distance to be the same apparent size as the host star. The bottom figure is moons that were scaled to the planet radius before distance was calculated.}
  \label{fig:PCcase2_3}
\end{figure*}
In Section~\ref{Data} we had filtered the population of known planets down to 38 well-characterized and self-consistent exoplanets to place moons around. In Section~\ref{Calculations}, we had used the following three approaches for placing exomoons around exoplanets:
\begin{enumerate}
    \item Lunar-radius exomoon placed at same distance as Earth-Moon distance
    \item Lunar-radius exomoon placed at requisite distance to have same apparent size as host star
    \item Exomoon scaled to Earth-Moon ratio and placed at requisite distance to have same apparent size as host star
\end{enumerate}

We then check these exomoons for two qualities; the first is that the exomoon does indeed have a possible orbit constrained by the Roche limit and the Hill sphere. For case 1, only a single exomoon meets this criterion; for case 2, 26 exomoons satisfy this criterion; for case 3, 29 exomoons satisfy this criterion. We then calculate the flux from each moon, and require this to exceed the flux the Earth receives from the full moon, $F_{\leftmoon}$. In all cases, so long as the moon had a potential orbit, the moon satisfied the minimum flux requirements. It remains to be seen, however, if there is such a thing as too bright a moon to allow for transmogrification.

Case 1 is the most intriguing, because the single planet that it finds capable of sustaining a qualifying moon is also included in cases 2 and 3, and is the exoplanet LHS 1140 b \citep{Dittmann2017}. This is a particularly compelling result as LHS 1140 b orbits a relatively nearby star ($\sim 10$ pc) within the liquid-water habitable zone, and so this system may be amenable to not just werewolves but to werehumans as well, while also being able to support an exomoon with all three cases we studied.

In Figure~\ref{fig:PCcase2_3} we display cases 2 and 3, plotting exomoon flux as a function of planetary semi-major axis. Planetary radii are represented by the size of the points, with the bottom left concentric circles showing radii of 0.5, 1, 1.5, and 2 Earth radii. The dashed bottom line at the bottom marks the flux the Earth receives from the full moon and the dotted line at the top represents typical solar flux. Host star masses are represented by color. We quickly see that all planets have semi-major axes less than 0.1 AU, likely reflecting biases in measurement completeness as close-in planets are easier to both detect and characterize. We also see that almost all host stars have masses less than our own Sun, and so represent a smaller, cooler stellar population.

Cases 2 and 3 yield many more exoplanets with exomoons providing flux above our threshold than case 1 did, although very few are close to the threshold, with many orders of magnitude brighter. If lycanthropic saturation is possible, many of these may not actually be suitable habitats to werewolves. From this, it does appear that with our more flexible criteria used in cases 2 and 3, roughly 70\% of exoplanets are capable of hosting an exomoon that would exceed the minimum flux needed to serve as a trigger for lycanthropy.

Following the previous investigation, we also conduct a somewhat more speculative investigation using the NExScI Exoplanet Archive's Planetary Systems Composite Parameters (PSCP) Table \citep{PSCT}, with the caveat that this may involve more inferred or inconsistent measurements. We then go through the same steps from Section~\ref{Data} in curating the data.

\begin{figure*}[!htb]
  \begin{center}
  \includegraphics[width=0.49\textwidth]{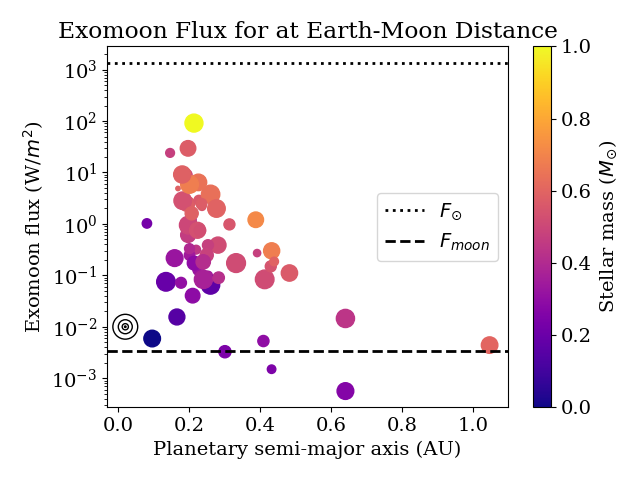}
   \includegraphics[width=0.49\textwidth]{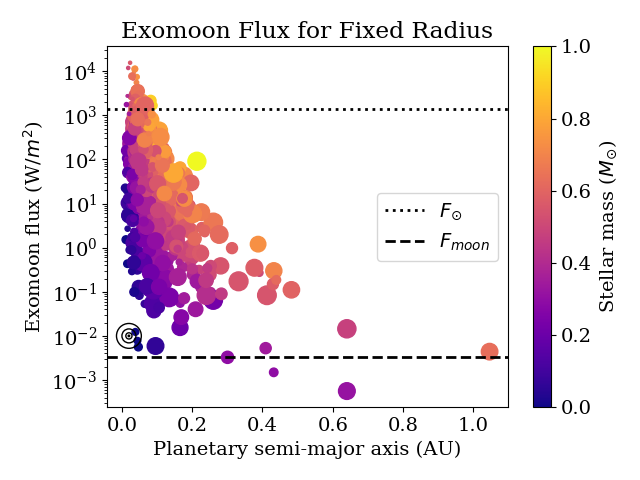}
   \includegraphics[width=0.49\textwidth]{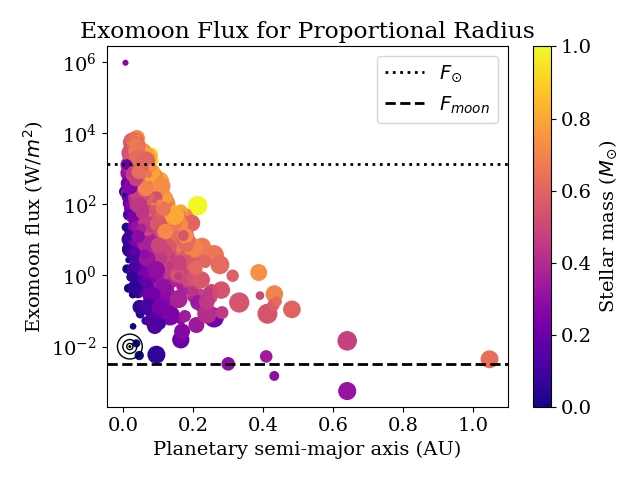}
  \end{center}
  \caption{The exomoon fluxes calculated for the vetted exoplanets from the Planetary Systems Composite Parameters table, with the planetary semi-major axis on the x-axis and the exomoon flux received by the planet on the y-axis. Points are scaled as a function of planetary radius, and the concentric circles in the bottom left represent radii of 0.5, 1, 1.5, and 2 Earth radii. Colors correspond to the host star's mass. The horizontal dashed line marks the flux the Earth receives from the full moon and the horizontal dotted line represents the flux the Earth receives from the Sun. The top-left and top-right are for when the moon's distance was determined to match apparent size of the host star based on a lunar-radius and a radius scaled to planet radius, respectively. The bottom figure is for a lunar-radius moon at 385,000 km.}
  \label{fig:PSCPcases}
\end{figure*}

Again, we begin with 5,005 exoplanets. When we filter out columns that contain missing data, this number drops to 4490 exoplanets. Including our radius cut for likely terrestrial planets reduces the sample to 1140, and our high-mass cut for terrestrial planets reduces this number slightly further to 1135 planets in the final sample. We again calculate exomoon fluxes as we did earlier, and display these results in Figure~\ref{fig:PSCPcases} using the same properties as in Figure~\ref{fig:PCcase2_3}.

We refer back to the three cases for the system architecture, and find that out of the 1135 planets, only 58 are able to host an exomoon at the same distance as the Earth-moon distance, almost all of which are as bright or brighter than $F_{\leftmoon}$.

Cases 2 and 3 represent 893 and 971 systems, respectively, and so roughly 80-85\% of terrestrial planets appear to be capable of hosting a moon capable of triggering lycanthropy. In both cases, as in Case 1, we see that almost all are brighter than the Earth's full moon, and most much brighter than that.

\section{Discussion}\label{Discussion}
\subsection{LHS 1140 b}
From our  study of well characterized exoplanets, the best candidate for an exoplanet to have an exomoon that gives rise to werewolves is LHS 1140b. This is not only because the planet is comparatively close and amenable to follow-up observation, but also because the hypothetical exomoon's flux would be very similar to the flux from our full Moon, only about twice as bright rather than the typical orders of magnitude brighter observed for other exoplanet systems and the planet itself orbits within its host star's habitable zone. Further supporting the case for potential habitability is the detection of water vapor in the atmosphere of LHS 1140 b \citep{Edwards2021}. While the equilibrium temperature of LHS 1140 b was reported as $\sim230$K \citep{Dittmann2017}, if there is a substantial greenhouse effect present from the planet's atmosphere, the climate at the planet's surface may be much closer to the rainy but temperate (275 to 300 K)\footnote{https://www.worldweatheronline.com/soho-weather-averages/westminster-greater-london/gb.aspx} climate of SoHo, itself known for its werewolves \citep{Zevon1978}.

LHS 1140 b also provides an example of an interesting demonstration on the impact of tidal effects on a star-planet-moon system. There is some suggestion that LHS 1140 b may represent a tidally-locked planet, in which case the full moon will only ever be observed on one side and somewhat more important impacts from only one side of the planet receiving sunlight \citep{Barnes2017, Spinelli2019}. This could mean a planet with one inhospitable side in perpetual sunlight and the other inhospitable side in frequent darkness and cold, and only a band of thriving life around the terminator, which would then be peppered with werewolf activity depending on the moon's phase \citep{Anders2019}.

As the exomoon in case 1 would have a period of approximately 10 days around the planet and LHS 1140 b itself has a period of around 25 days around its host star, a tidally locked planet would also result in an orbit of the exomoon that undergoes significant change, as the slower planetary rotation could cause the moon to spiral inward due to tidal interactions between the planet and moon.

\subsection{Population-level Effects}

For large scale characterization there is very little support in the literature for understanding if future work should prioritize exomoon distance, exomoon radius, or exomoon/exoplanet ratio when attempting to identify the systems that are most capable of supporting werewolves. It is unclear if luminosity does indeed have a saturation point above which transformation cannot occur. Additionally, the potential impact of stellar properties has gone almost entirely unaddressed, as the sun has been a fixed variable in past analyses of werewolf environments, with the larger focus on the "inconstant moon, [t]]hat monthly changes in her circle orb" \citep{Shakespeare1597}.

The answers to these questions are largely absent from the literature, likely as this sort of research has been conducted in the dark for the last several decades. Only Dr. Calen Henderson, formerly at NASA, has provided any explanation \citep{Hodges1980,Queen1980}:
\begin{quote}
This is an active area of top-secret research, but we think our initial hunch --- that the most critical trait is the total amount of incident flux --- may yet turn out to be correct, at least to first-order.

Further research is required, but this would make the "optimal" were-moon, so to speak, depend on a complex combination of the lunar albedo, the were-moon's orbital semi-major axis, the orbital semi-major axis of the were-planet+moon system, and the SED of the host star(s).\footnote{personal communication with the author, March 29, 2022}
\end{quote}

The interest in SEDs of host stars is particularly notable as there has been some work into the other impacts that different stellar types can have on other biological functions, as evidenced by the photonucleic effect discussed in \citet{Maggin1999}:
\begin{quote}
    The photonucleic effect is a very specialized astronomical phenomenon. It takes place when an object native to the influence of a red giant star enters the influence of a small G-type star...
    The first thing that happens under the photonucleic effect is that the nucleus of each of an organic creature's cells grows a carapace—a temporary shell that shields it from external harm. This happens very quickly, from the outside of the organism - the cells in direct contact with yellow starlight—to the most internal cells of the organs, in less than the time it takes to draw a breath.
\end{quote}

It is important to note an important distinction in that \citeauthor{Maggin1999} is discussing a phenomenon that occurs when a given living organism that is normally exposed to one kind of light, such as that from a red dwarf, is then exposed to a different star, such as our Sun. As such, this is more analogous to the transformations that occur in a single lycanthropic organism when transitioning between sunlight and moonlight than it is to the study of comparative behavior of unrelated organisms living exclusively around different types of stars.

\subsection{WFIRST}
Dr. Henderson's interest in SED of host stars corresponds quite interestingly to the results we found for our stellar populations in both Figure~\ref{fig:PCcase2_3} and Figure~\ref{fig:PSCPcases}. Even if we work within a framework that includes some sort of saturation point, the exoplanets that seemed most amenable to hosting exomoons that support lycanthropy are in orbit around M-dwarfs. At those much cooler temperatures, it would not be unreasonable to expect that this werewolf behavior would need to be capable of being driven by an entirely different part of the spectrum.

In order to study this, we suggest a targeted space-based mission we call the Werewolves From Infrared Radiation and Spectral-typing Telescope, or WFIRST for short. We believe that this mission could benefit from significant cost savings by using this name now that the Nancy Grace Roman Space Telescope (RST) has dropped usage of this name \citep{NASA2020}, as there would likely be significant amounts of stationary and graphic design work that could be reused, reducing mission overhead and allowing a greater focus on the more important matter of werewolves.

\subsection{Future Steps}
This work has relied on a relatively simple framework for the systems examined, consisting of a host star, an exoplanet, and an exomoon. Future work should address these tidal influences, as this is known to have a significant impact on the potential for exomoons to survive over longer time-scales when looking at M-dwarfs \citep{Zollinger2017, Martinez2019}.

Another more dynamic, but still likely, system would be that of an exoplanet with more than one moon. In these cases, it may be possible for more than one full moon to be present at the same time. It is not clear initially if two full moons would represent a linear increase in lycanthropic behavior or an exponential one; we refer to these possibilities as a double werewolf and a squarewolf respectively. More complex systems also would make a three werewolf moon a possibility depending on the shape of this function.

Along a similar vein, we have constrained ourselves to standard full moons, and have not looked at the impact that new moons or eclipses could have on werewolves, or the interplay between sunlight and moonlight that occurs at twilight or the breaking dawn, or in polar regions where a midnight sun is present. The impacts these have may be life or death questions for any werewolves in this environments, although we are not yet on either Team with respect to their significance.

The most exotic scenario we note for future exploration would be a system in which the illumination of a full moon is provided by an object that is not, itself, gravitationally bound to the planet but does pass the planet close enough to have the same influence as an exomoon. This could be categorized as a particular flavor of the famed Steppenwolf planet type that has been proposed by \citet{Abbot2011}. However, as that generally represents rogue planets that are large enough to maintain an atmosphere and be independently habitable, we suggest that a somewhat more generalized term should be used. A very fitting description of this approximately lunar-mass, free-floating objects would be solivagant planetary-mass objects (SPlaMOs) \citep{Christiansen2020}. These SPlaMOs have the potential to create outbursts of lyncanthropy on planets that wouldn't observe periodic occurrences linked to an exomoon orbit.

In the broader field of exocryptozoology, we consider this work characterizing werewolves in space to be the next step in understanding the prevalence of Universal monsters. This follows the ground work, or rather space work, done in \citet{Gunther2020} to prepare for future opportunities to detect vampires in space (while the work in \citep{Kane2014} is of high caliber and worthy of note, it should be stated that its subject was not Universal). There are, however, very tantalizing new directions to go in by better understanding the pyramidal-shaped Ahuna Mons on Ceres \citep{Raponi2016, Freund1932} and looking for creatures from the Lagoon Nebula \citep{Arnold1954}.

\section{Summary}\label{Summary}
In this work, we have established the thousands-of-years-long chronology that has inevitably intertwined astronomy and lycanthropy. We then establish some criteria that a moon must meet in order to be able to trigger transformation, and identify the potentially habitable zone planet LHS 1140 b as a leading contender for hosting an exomoon and being inhabited by werewolves.

\section{Acknowledgements}
The authors wish to thank Calen Henderson's input in determining the scope of this work and developing a future support mission, Adiv Paradise's willingness to respond to social media posts vaguely asking for input relating to these topics and insightful comments on the methodology and tidal influences, and Savannah Jacklin's detailed wordsmithing. We also thank them for the words SPlaMOs, exocryptozoology, and squarewolf, respectively.

This research has made use of the NASA Exoplanet Archive, which is operated by the California Institute of Technology, under contract with the National Aeronautics and Space Administration under the Exoplanet Exploration Program.

This research made use of Astropy,\footnote{http://www.astropy.org} a community-developed core Python package for Astronomy \citep{astropy:2013, astropy:2018}.

This research has made use of NASA’s Astrophysics Data System.

Software: astropy \citep{astropy:2018}, matplotlib \citep{Hunter2007}, numpy \citep{Oliphant2006}, pandas \citep{Mckinney2011}


\bibliographystyle{apalike}
\bibliography{main}


\end{multicols}

\end{document}